\documentclass[aps,prd,twocolumn,superscriptaddress]{revtex4-1}

\usepackage{graphicx}
\usepackage{color}
\usepackage{amsfonts}
\newif\ifoneauthor
\oneauthortrue
\newcommand{\we}[0]{\ifoneauthor I \else we \fi}
\newcommand{\are}[0]{\ifoneauthor am \else are \fi}
\newcommand{\We}[0]{\ifoneauthor I \else We \fi}
\newcommand{\omg}[0]{{{\omega}}}

\newcommand{\del}[0]{{}_{{}^\triangle}\!}
\newcommand{\w}[0]{{\rm w}}
\newcommand{\e}[0]{{\rm e}}

\newcommand{\eq}[1]{equation \ref{#1}}
\begin{document}

\title{Photo-Thermal Transfer Function of Dielectric Mirrors for Precision Measurements}


\author{Stefan W. Ballmer}
\email[]{sballmer@syr.edu}
\affiliation{Department of Physics, Syracuse University, NY 13244, USA}



\date{\today}

\begin{abstract}
The photo-thermal transfer function from absorbed power incident on a dielectric mirror to the effective mirror position is calculated using the coating design as input. The effect is found to change in amplitude and sign for frequencies corresponding to diffusion length comparable to the coating thickness. Transfer functions are calculated for the $Ti$-doped ${\rm Ta_2O_5:SiO_2}$ coating used in Advanced LIGO and for a crystalline ${\rm Al_xGa_{1-x}As}$ coating.  The shape of the transfer function at high frequencies is shown to be a sensitive indicator of the effective absorption depth, providing a potentially powerful tool to distinguish coating-internal absorption from surface contamination related absorption.
The sign change of the photo-thermal effect could also be useful to stabilize radiation pressure-based opto-mechanical systems. High frequency corrections to the previously published thermo-optic noise estimates are also provided.
Finally, estimating the quality of the thermo-optic noise cancellation occurring in fine-tuned ${\rm Al_xGa_{1-x}As}$ coatings requires the detailed heat flow analysis done in this paper.
\end{abstract}


\maketitle

\section{Introduction}
The photo-thermal effect is the coupling from fluctuations in absorbed power incident on a mirror to the effective mirror position. It is important for a wide range of applications involving varying amounts of power incident on a mirror. Examples range from a source of noise coupling in gravitational-wave interferometers to an important feed-back path in many types of mico-electromechanical systems. Additionally, the photo-thermal effect is closely related to the thermo-optic noise in mirror coatings, which is one of the limiting noise sources for upgrades to the gravitational-wave interferometers currently under construction (Advanced LIGO \cite{Harry2010}, 
Advanced Virgo \cite{TheVirgo:2014hva} and Kagra \cite{Somiya:2011np}).
The importance of the effect for gravitational wave detectors has driven a theoretical \cite{Braginsky1999, Braginsky2000, Levin1998, PhysRevD.63.082003, Levin2008, Levin2009, Evans2008, PhysRevD.90.043013} and experimental \cite{Rao2003, Harry:2001iw, PhysRevLett.89.237402, Harry:06, Harry2010, Cole2013} interest in understanding and improving the fundamental thermal noise of optical elements.

The photo-thermal transfer function takes a simple form at frequencies for which the diffusion length $d_{\rm diff}$  is small compared to the transverse dimension of the beam spot, but large compared to the coating thickness $d_{\rm coat}$. Both the thermal diffusion and the elasticity problem become one-dimensional, and the resulting mirror surface displacement is the integral of the deposited heat  (\eq{eq:simple} below).
In \cite{PhysRevD.63.082003} Cerdonio et al. explored corrections to this simple picture that arise due to transverse diffusion. Their result predicted a decrease in the photo-thermal response for frequencies corresponding to diffusion length comparable to the beam spot (\eq{eq:Cerdonio} below). This result was later experimentally confirmed by De Rosa et al. \cite{PhysRevLett.89.237402}, measuring the coupling up to $200~{\rm Hz}$. Both papers however assume that the diffusion length is much bigger than the coating thickness.

The dielectric stack of the mirror coating affects both the heat diffusion at higher frequencies and the local coupling of temperature to the total reflected phase of the coating, which is the quantity that determines the mirror position read-out. The latter was discussed in detail in \cite{Evans2008} in the context of exploring thermo-optic noise. In particular we found that, compared to substrate heating, heating of the first couple of dielectric layers has the opposite effect on the mirror position read-out. This is caused by the change in optical thickness of the dielectric layers. However \cite{Evans2008}  did not discuss the implications for the photo-thermal  transfer function, which is explored in this paper.

The rest of the paper is structured as follows: Section \ref{PTc} revisits the previously published properties of the photo-thermal effect. In section \ref{Derive} the photo-thermal transfer function is calculated taking into account the coating structure. Section \ref{Discuss} applies the result to both the $Ti$-doped ${\rm Ta_2O_5:SiO_2}$ coating used in Advanced LIGO and for a crystalline ${\rm Al_xGa_{1-x}As}$ coating. In section \ref{TOnoise} the implications for thermo-optic noise are discussed.

\section{Substrate Photo-thermal Coupling}
\label{PTc}
Throughout this paper we will work in the complex Fourier domain, i.e. observed quantities like the effective mirror displacement $\del z$ are complex and obey the relation $\del z^{*}(\omg) = \del z(-\omg)$.

We are interested in the photo-thermal transfer function from the absorbed surface intensity $j$ to effective mirror displacement $\del z$, i.e. the mirror position as read out by a laser beam.
In the limit where the coating thickness $d_{\rm coat}$ is negligible compared to the diffusion length $d_{\rm diff}$ and transverse diffusion is irrelevant,  $d_{\rm coat} \!\!\ll\!\! d_{\rm diff} \!\!\ll\!\! \w$, with $\w$ is the Gaussian beam radius, we can solve the one-dimensional heat diffusion equation
\begin{equation}
\partial_z j(z) + \rho C \dot{T}(z) =0
\end{equation}
with the boundary conditions $j(\infty)=0$, and $j(0)=j$, the absorbed surface intensity. The mirror displacement $\del z$, and therefore the photo-thermal transfer function take the form 
\begin{equation}
\label{eq:firstsimple}
\del z = \bar{\alpha} \int_0^\infty \!\!\!\!\!\!T dz = \bar{\alpha} \frac{j}{i \omg \rho C} .
\label{eq:simple}
\end{equation}
Here $d_{\rm diff}=\sqrt{\kappa/(\rho C \omg)}$ is the diffusion length in the substrate, with $\kappa$, $C$ and $\rho$ the thermal conductivity, heat capacity and density of the material. $\omg$ and  $j$ are the observation frequency and the absorbed surface intensity. Finally, $\bar{\alpha}=2(1+\sigma) \alpha$ is the effective expansion coefficient under the mechanical constraint that the heated spot is part of a much larger optic, see for instance \cite{Evans2008}, equation A1, or \cite{PhysRevD.70.082003}. $\alpha$ and  $\sigma$ are the regular linear expansion coefficient and the Poisson ratio.

Cerdonio et al. \cite{PhysRevD.63.082003} expanded this to include the effect of transverse diffusion, and found that \eq{eq:firstsimple} needs an additional multiplicative correction factor
\begin{equation}
\label{eq:Cerdonio}
I(\Omega) = \frac{1}{\pi} \int\limits_0^\infty du \int\limits_{-\infty}^\infty dv \frac{u^2 \e^{-u^2/2}}{(u^2+v^2)\left(1+\frac{(u^2+v^2)}{i \Omega} \right) } ,
\end{equation}
with $\Omega = \omg {C \rho \w^2}/{(2 \kappa)}$. As expected, for $\Omega \gg 1$, the correction factor becomes $I(\Omega) \approx 1$. The magnitude of \eq{eq:Cerdonio} was experimentally confirmed by De Rosa et. al. \cite{PhysRevLett.89.237402}. 

\section{Coating Correction}
\label{Derive}
To calculate the effect of a dielectric coating on the photo-thermal transfer function
we first have to find the response of the coating reflected phase to temperature fluctuations at each layer,  and then solve the heat diffusion equation. If we include the detailed coating structure and transverse diffusion we will find the full photo-thermal transfer function. However, dielectric optical coatings require that the spot size $\w$ is much bigger than the coating thickness $d_{\rm coat}$ - otherwise the plane wave approximation inside the coating is not justified. Since the frequency dependent diffusion length $d_{\rm diff}$ is the relevant scale parameter for both coating effects and transverse diffusion effects, and since 
the limit  $d_{\rm coat} \!\!\ll\!\! d_{\rm diff} \!\!\ll\!\! \w$ results in the simple expression from equation \ref{eq:simple}, the corrections due to transverse diffusion and the corrections due to the coating structure will never be big at the same time. Furthermore, expanding the heat diffusion equation as Taylor series shows that to first order one can simply calculate both effects as multiplicative corrections.
Thus for the following calculation \we can ignore any transverse heat diffusion without loss of generality.

First I calculate the response of the coating reflected phase to temperature fluctuations at each coating layer. Following the notation of \cite{Evans2008}, the change $\del \phi_k$ in the optical round trip phase in coating layer $k$ due to temperature fluctuations is given by the following integral across layer $k$:
\begin{equation}
\del \phi_k = \frac{4 \pi}{\lambda_0}  \int\limits_{k}  (\beta_k + \bar{\alpha}_k n_k) T(z) dz ,
\end{equation}
where $\bar{\alpha}_k$ is the effective expansion coefficient under the mechanical constraint from the coating being attached to a substrate, as discussed in \cite{Evans2008}, equation A1, or \cite{PhysRevD.70.082003}:
\begin{equation}
\bar{\alpha}_k =\alpha_k \frac{1+\sigma_s}{1-\sigma_k} \left[ 
\frac{1+\sigma_k}{1+\sigma_s} + (1-2 \sigma_s) \frac{E_k}{E_s} \right] .
\end{equation}
$E_k$, $E_s$, $\sigma_k$ and $\sigma_s$ are the Young's moduli and Poisson ratio for layer $k$ and the substrate. If the coating layers have similar elastic properties this becomes $\bar{\alpha}_k~\approx~2(1~+~\sigma)~\alpha_k$.

The coupling of $\del \phi_k$ to the phase of the light reflected of the coating $\del \phi_{\rm c}$ is given by
\begin{equation}
\label{eq:dphidphi}
\frac{\partial \phi_{\rm c}}{\partial \phi_k} = {\rm Im}{\frac{1}{r}\frac{\partial r}{\partial \phi_{k}}},
\end{equation}
where $r$ is the complex field reflectivity of the coating. In \cite{Evans2008} a recursive expression for these partial derivatives is given, and they are shown to be negative for quarter wavelength coatings. Appendix~\ref{App:Coat} gives an alternate approach to calculating them.

Additionally all layers and the substrate also contribute to the total expansion of the mirror. If \we set $\frac{\partial \phi_{\rm c}}{\partial \phi_s}=0$ and include the substrate in the summation, the total change of the coating reflected phase $\del \phi_{\rm c}$ becomes
\begin{equation}
\label{eq:dphic}
\del \phi_{\rm c} = \frac{4 \pi}{\lambda_0}  \int_0^{\infty}   \left[ \frac{\partial \phi_{\rm c}}{\partial \phi_k} (\beta_k \!+\! \bar{\alpha}_k n_k) 
\!+\!  \bar{\alpha}_k  \right]  T(z) dz ,
\end{equation}
where the material parameters in the brackets are evaluated for the layer $k$ that contains the volume element at depth $z$.

Next, \we solve the one-dimensional heat diffusion equation across the coating.
For simplifying the derivation \we assume that all heat is deposited on the first interface layer. This is not necessarily a good assumption as the field penetrates a couple layers into the coating. Extending the analysis to bulk absorption is briefly discussed at the end of  appendix \ref{App:CoatDiffuse}, and results in a small change to equations \ref{eq:layersolution}, \ref{eq:Tidi} and \ref{eq:Tsds}.
\We now define $\xi_k=\sqrt{{i \omg C_k \rho_k}/{\kappa_k}}$ for every layer $k$. Inside this layer the heat diffusion equation is 
$\xi_k^2 T = T'' $, where the notation $' = \partial_z$ is used. This has the solution
\begin{equation}
\label{eq:layersolution}
T(z) = T_R \e^{-\xi (z-z_0)} +  T_L \e^{\xi (z-z_0)} ,
\end{equation}
where $T_R$ and $T_L$ are the right-propagating and left-propagating mode amplitudes at $z=z_0$.
The solution for the temperature profile across the whole coating can now be found by matching the boundary conditions. Specifically,  $T$ and $j=-\kappa_k \nabla T$ are continuous across coating boundaries, and $j$ at the surface is equal to the external heating power. A solution is given in appendix \ref{App:CoatDiffuse}.

To evaluate equation~\ref{eq:dphic} the temperature integral across every layer $k$ is needed. It can be expressed as a function of the temperature in the middle of the layer:
\begin{equation}
\label{eq:Tidi}
\bar{T}_k d_k := \int_{-d_k/2}^{d_k/2} T(z) dz =\frac{2}{\xi_k} \sinh (\frac{\xi_k}{2} d_k )  (T_R + T_L)_{\rm middle} .
\end{equation}
Similarly for the substrate \we can define 
\begin{equation}
\label{eq:Tsds}
\bar{T}_s d_s := \int_{0}^{\infty} T(z) dz = \frac{T_{R,s}}{\xi_s} ,
\end{equation}
where $T_{R,s}$ is the temperature at the coating-substrate interface.
The total change of the effective mirror position thus becomes
\begin{equation}
\label{eq:dphic2}
\del z =  \sum_{k}   \left[ \frac{\partial \phi_{\rm c}}{\partial \phi_k} (\beta_k \!+\! \bar{\alpha}_k n_k) 
\!+\!  \bar{\alpha}_k  \right]  \bar{T}_k d_k ,
\end{equation}
where the sum goes over all layers plus the substrate, and \we used $\del \phi_{\rm c} = {4 \pi} \del z/ {\lambda_0}$. The bracket in \eq{eq:dphic2} is negative and relatively large for the first few layers of the coating, and becomes positive closer to the substrate.

Now all the pieces for calculating the photo-thermal transfer function are in place. Since \we am working in the Fourier domain, \we will evaluate it one frequency at a time: (i) Given the surface heating $j$, \we calculate the temperature profile across the coating. In particular \we \are interested in the temperatures in the middle of the coating layers, given by equation \ref{eq:tempprofile}. (ii) \We calculate the partial derivatives $\frac{\partial \phi_{\rm c}}{\partial \phi_k}$ using equation \ref{eq:parder} to get the sensitivity of the coating to round trip phase changes in each layer. (iii) \We can now evaluate equation \ref{eq:dphic2} to find the effective mirror displacement $\del z$. The ratio $\del z/j$ is the photo-thermal transfer function. 

At low frequencies ($d_{\rm coat} \!\!\ll\!\! d_{\rm diff}$) the temperature fluctuations will reach far into the substrate. Thus only the substrate term in equation \ref{eq:dphic2} will be relevant, and \eq{eq:firstsimple} is recovered. On the other hand, for frequencies with $d_{\rm diff}$ smaller than $d_{\rm coat}$, the negative bracket in \eq{eq:dphic2} for the first few coating layers results in an enhancement and a sign change of the transfer function, as we will see on concrete examples in the next section.

\section{Discussion and Implications}
\label{Discuss}
\begin{figure*}[htb]
  \centering
  \includegraphics[width=2 \columnwidth]{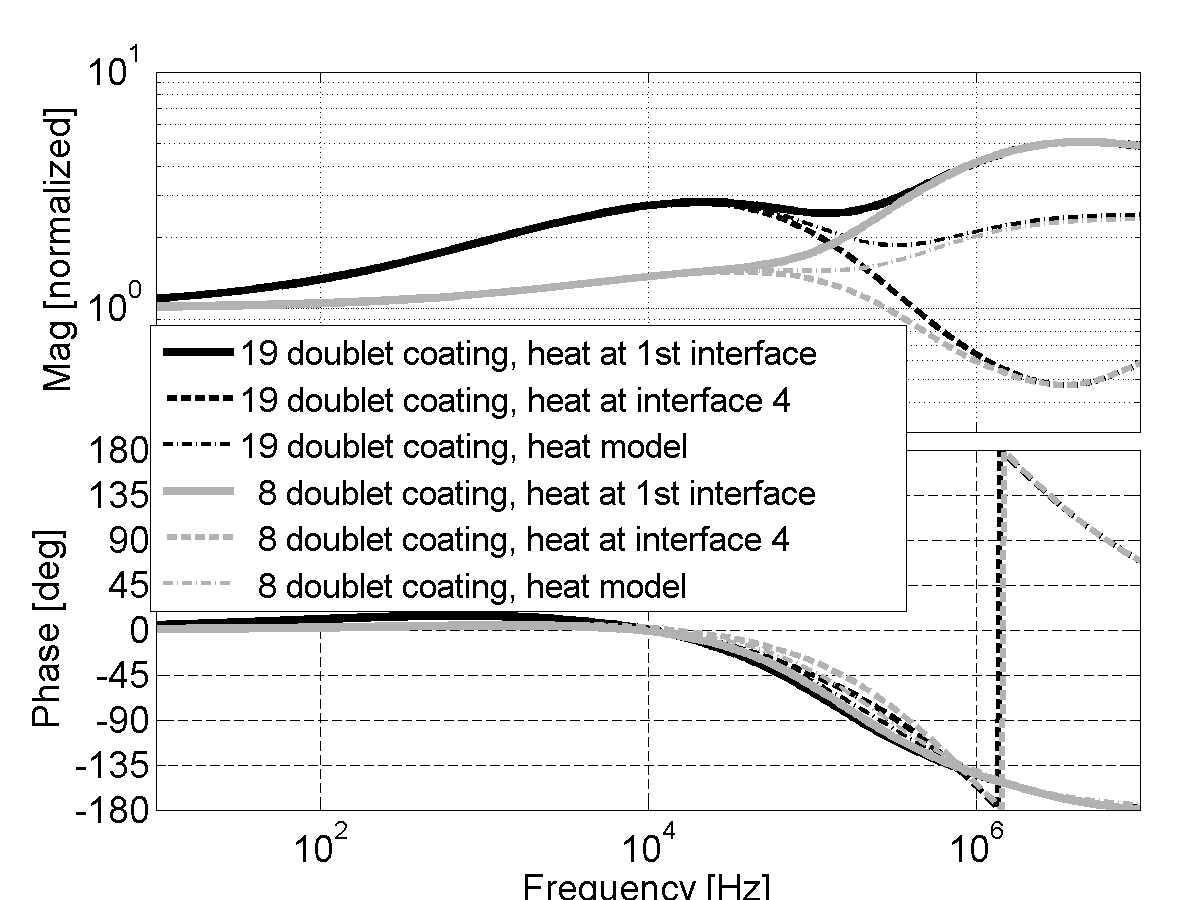}
  \caption{A bode plot of the photo-thermal transfer function correction factor for a $\rm Ta_2O_5:SiO_2$ coating. The black (gray) traces are for a 19-doublet (8-doublet) coating, corresponding to a Advanced LIGO end (input) test mass coating. For the solid traces the heat was deposited at the front surface of the coating. For the dashed traces it was deposited at the fourth interface layer, at a depth of $0.68~{\mu{\rm m}}$. Finally, for the dash-dotted traces, the power was deposited in the coating according to the optical power present in each layer. At high frequencies the transfer function strongly depends on heat deposition depth, which in turn can be explotited to measure the absorption depth (see text). To get the full transfer function multiply with \eq{eq:firstsimple}. The calculation is based on the parameters from table \ref{SiO2}.
}
\label{fig1}
\end{figure*}
First \we evaluate the photo-thermal transfer function for two coatings of interest for the gravitational-wave community. 
\We start with a quarter-wave $\rm Ta_2O_5\!\!:\!\!SiO_2$ coating, heated at the coating surface. The Advanced LIGO end and input test masses are coated with a titanium-doped $\rm Ta_2O_5\!\!:\!\!SiO_2$ coating with 19 and 8 doublet layers respectively. For clarity \we divide out the naive expectation from \eq{eq:firstsimple} and ignore the beam spot size dependence from equation \ref{eq:Cerdonio}. Note that for the case of Advanced LIGO ($\w=6~{\rm cm}$) this is a good approximation. Even at the lower edge of the Advanced LIGO observation band ($10~{\rm Hz}$) the normalized frequency $\Omega$ from equation \ref{eq:Cerdonio} is already $1.3\cdot 10^5$ and transverse diffusion is not important.
The correction factor arising from the coating structure is shown in figure \ref{fig1} for both coatings (black and gray solid traces).  As expected a gradual sign change and increase in magnitude is occurring around about $100~{\rm kHz}$. The correction factor however has a tail that extends to relatively low frequencies, reaching $3~{\rm dB}$ at $160~{\rm Hz}$ and $6~{\rm dB}$ at $1~{\rm kHz}$ for the 19-doublet coating. Note that the high frequency feature significantly depends on the depth at which the heat is deposited, while the same is not true for the low frequency tail. To illustrate this point figure \ref{fig1} shows two additional traces for each coating. The dashed traces correspond to transfer functions for which the heat was deposited at a depth of $0.68~{\mu{\rm m}}$, that is at the 4th interface layer (beginning of the 2nd high-index layer). The dash-dotted traces correspond to a model in which the power absorbed in each layer is proportional to the optical power circulation at that depth. This is a realistic absorption model if the absorption is not dominated by surface contamination. If all the heat is deposited at the 6th interface layer or deeper (not shown in figure \ref{fig1}), the sign change or phase wrapping at higher frequencies that is seen in all traces in figure \ref{fig1} will disappears. The photo-thermal effect is then dominated by simple material expansion at all frequencies.

We therefore see that at high frequencies the transfer function is a sensitive function of the heat deposition depth.
This effect can be exploited to measure the depth at which the optical absorption in the coating occurs. This approach could be a powerful diagnostic tool to distinguish intrinsic absorption inside the coating from contamination on the coating surface.

\begin{figure}[htb]
  \centering
  \includegraphics[width=\columnwidth]{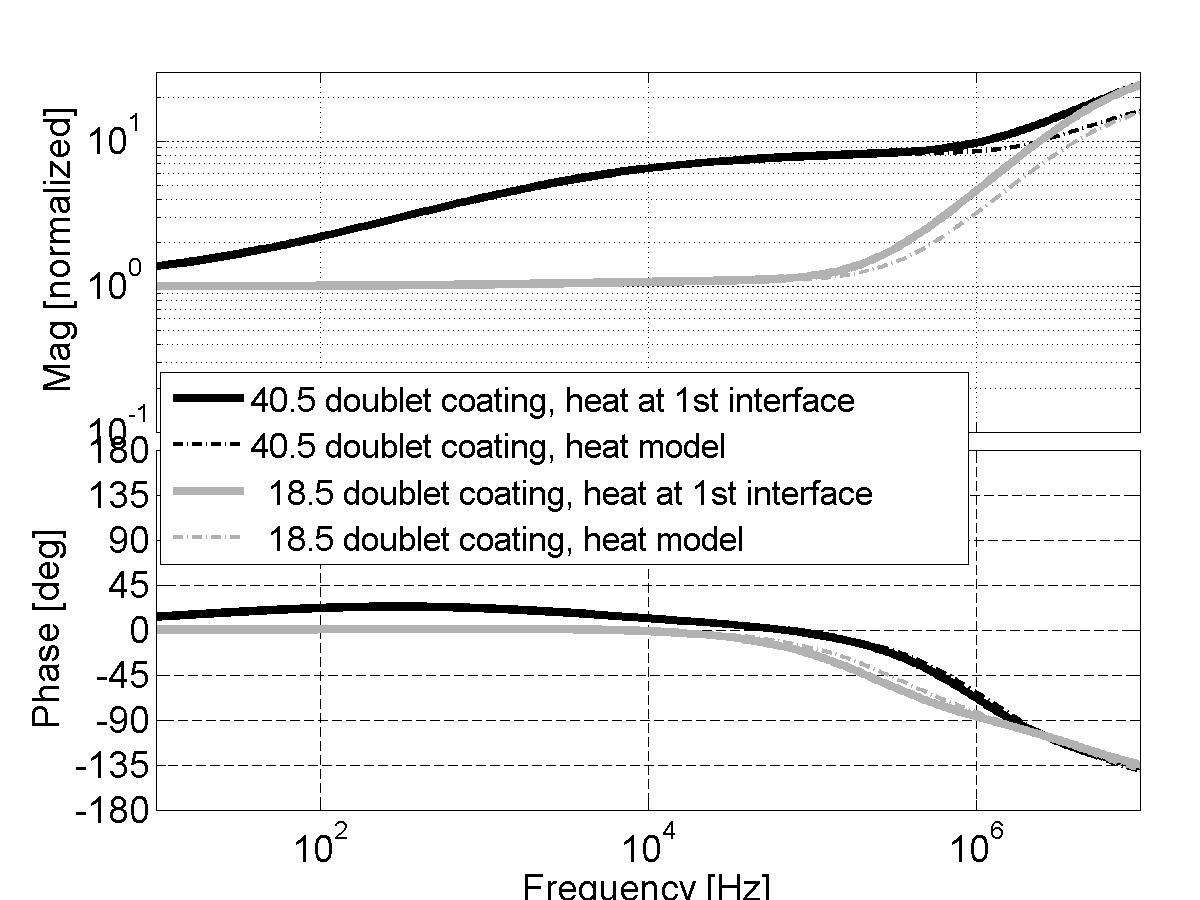}
  \caption{A bode plot of the photo-thermal transfer function correction factor for a
$\rm GaAs:Al_{0.92}Ga_{0.08}As$ coating with 40.5 and 18.5 $\lambda/4$ doublets (black and gray). 
For the solid traces the heat was deposited at the front surface of the coating. For the dash-dotted traces, the power was deposited in the coating according to the optical power present in each layer. To get the full transfer function multiply with \eq{eq:firstsimple}. The calculation is based on the parameters from table \ref{AlGaAs}.
}
\label{fig2}
\end{figure}
Next, figure \ref{fig2} shows the photo-thermal transfer function correction factor for the crystalline $\rm GaAs\!:\!Al_{0.92}Ga_{0.08}As$ coating discussed in \cite{Cole2013}. Shown in black and gray are traces for coatings with 40.5 and 18.5 $\lambda/4$ doublets respectively. They correspond to power reflectivities of $(1-2.5{\rm ppm})$ and $0.9976$.
Due to the higher heat conductivity of the crystalline coating the transfer function is much less dependent on the absorption depth. The flatness of the 18.5 doublet photo-thermal transfer function implies that the coating has no influence on the total photo-thermal effect, i.e. the substrate photo-thermal effect acquires no correction due to the coating. This is a sign of the cancellation effect between thermal expansion and index of refraction change that naturally occurs for this particular coating. As shown below this leads to a significant thermo-optic noise cancellation.

There are several implications worth discussing here. First this calculation predicts a small change in the expected intensity noise coupling in the observation band of gravitational-wave detectors. The Advanced LIGO mirrors are expected to have a coating absorption coefficient of less than $1~{\rm ppm}$, which should keep photo-thermal shot noise below the design quantum noise. The effect would be more important for any compensation system for thermal lensing that relies of projecting a heating pattern on to the surface of a test optic. This is currently not planned for Advanced LIGO exactly because of the photo-thermal effect \cite{BallmerThesis}. 
Comparing  $\rm GaAs:Al_{0.92}Ga_{0.08}As$ coatings to  $\rm Ta_2O_5:SiO_2$  coatings, the influence of the coating onto the photo-thermal transfer function is nominally slightly larger in the gravitational wave observation band. However the higher thermal conductivity tends to equalize the temperature fluctuations across the whole coating, making it easier to design a coating for which the photo-thermal effect cancels across a wide band \cite{TaraThesis,RanaTBD}.

The photo-thermal effect is also important for any opto-mechanical feed-back system, as it tends to dominate over the radiation pressure at higher frequencies. Due to the cavity response time,  radiation pressure based single carrier optical spring systems are either statically or dynamically unstable. A second optical carrier is needed to get stable optical feed-back \cite{Corbitt07,Perreca14}. The photo-thermal effect due to residual absorption will slightly change the phase of the optical spring. Indeed, the first-order effect given in \eq{eq:firstsimple} will always drive the optical spring towards instability. If however an optical spring has a resonance frequency high enough that the photo-thermal effect changes sign, in the case of figure \ref{fig1} above about 100~kHz, the photo-thermal effect will tend to stabilize the optical spring. The additional photo-thermal feed-back can indeed overcome the feed-back delay due the cavity response time, and lead to a cavity self-locking effect. This holds even for a single-carrier optical spring.

\section{Thermo-Optic Noise}
\label{TOnoise}
\begin{figure}[htb]
  \centering
  \includegraphics[width=\columnwidth]{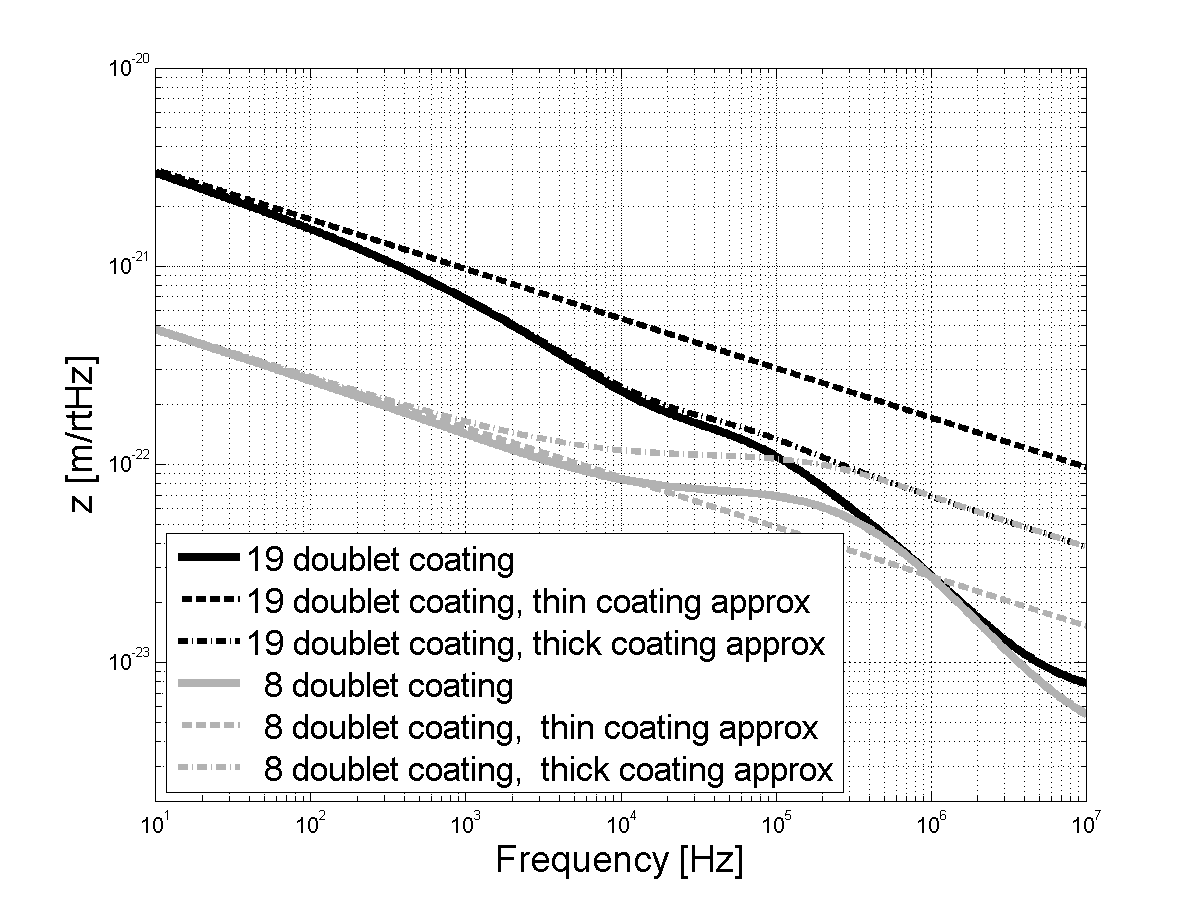}
  \caption{Thermo-optic noise of a $\rm Ta_2O_5:SiO_2$ coating with 19 and 8 doublets (black and gray). The solid trace is based on the full heat flow calculation in the coating. The dashed and dash-dotted traces are the thin and thick coating approximations discussed in \cite{Evans2008}. The calculation is based on the parameters from table \ref{SiO2} and a beam spot size of $\w=6~{\rm cm}$.
}
\label{fig3}
\end{figure}
\begin{figure}[htb]
  \centering
  \includegraphics[width=\columnwidth]{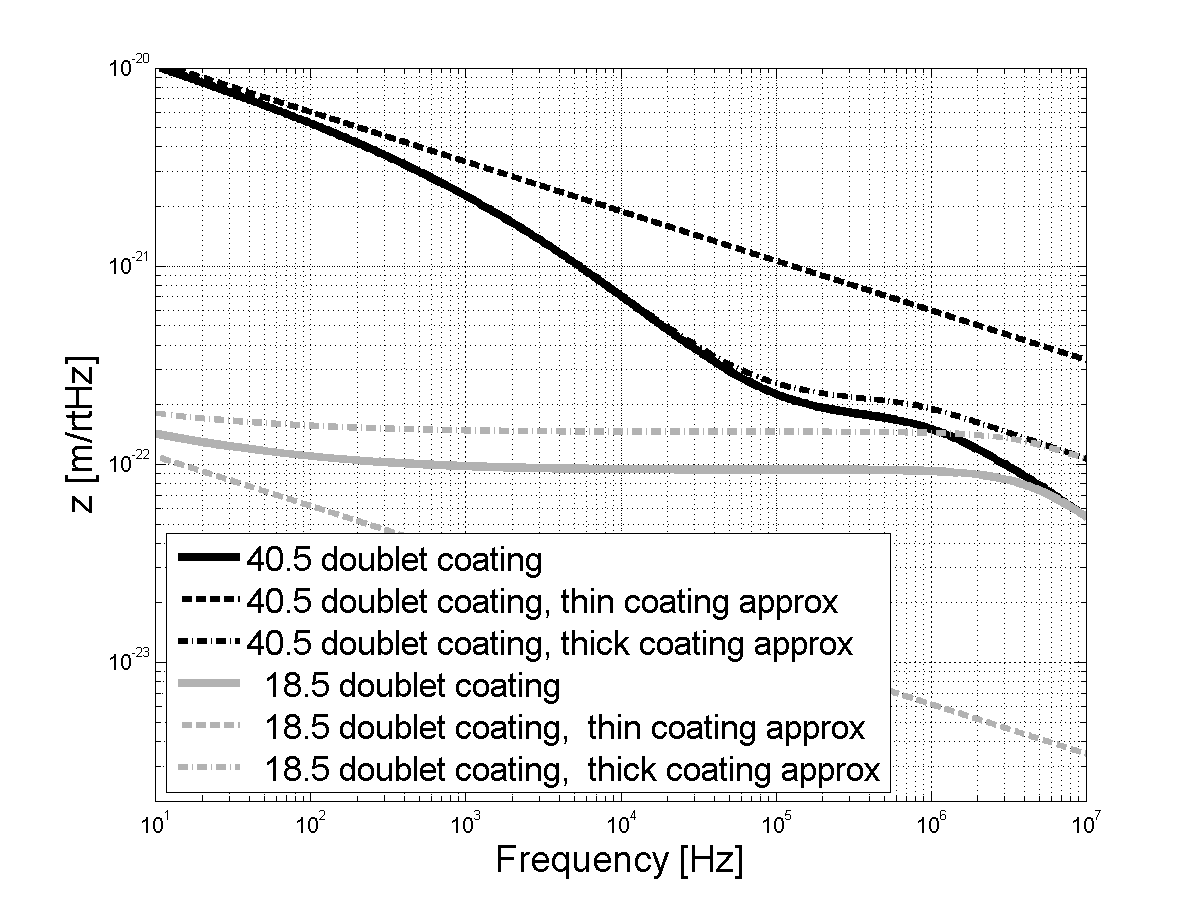}
  \caption{Thermo-optic noise of a $\rm GaAs\!:\!Al_{0.92}Ga_{0.08}As$ coating with 40.5 and 18.5 $\lambda/4$ doublets (black and gray), corresponding to power reflectivities of $(1-2.5{\rm ppm})$ and $0.9976$. The solid trace is based on the full heat flow calculation in the coating. The dashed and dash-dotted traces are again the thin and thick coating approximations discussed in \cite{Evans2008}, applied to the crystalline coating. A cancellation of the noise coupling naturally occurs for a 18.5 layer doublet, but can be engineered for higher reflectivity coatings by deviating from the simple $\lambda/4$ structure \cite{TaraThesis}. The calculation is based on the parameters from table \ref{AlGaAs} and a beam spot size of $\w=6~{\rm cm}$.
}
\label{fig4}
\end{figure}

In his 2008 paper \cite{Levin2008} Levin applied the fluctuation-dissipation theorem to thermo-refractive noise calculation. He was interested in the thermal noise seen by one specific degree of freedom, such as the mirror displacement $\del z$ read out by a laser beam. $\del z$ is a linear function of the temperature field in the optic: 
\begin{equation}
\del z =  \int dV  q(r,z) \delta T(r,z) .
\end{equation}
Levin showed that the thermal noise seen by this degree of freedom is proportional to the dissipated power per cycle if we inject an entropy with the same spatial profile $q(r,z)$. For the calculation of the photo-thermal transfer function we had to solve the heat diffusion equation and found the full heat flow field in the optic as a by-product (appendix \ref{App:CoatDiffuse}). This directly permits calculating the dissipated power. Taking into account the full details of the coating we can thus use the same approach to calculate the thermo-optic noise, that is the coupling of temperature fluctuations due to the combined effect of thermal expansion (thermo-elastic noise) and change in index of refraction (thermo-refractive noise). More details on the thermal noise calculation is given in appendix \ref{App:Noise}.

In \cite{Evans2008} we gave approximations for both a thin-coating limit and a thick coating extension that describes the correlation of thermo-elastic and thermo-refractive noise as a function of coating thickness. This thick coating extension is a good approximation for the observation band of Advanced LIGO. We did however use coating-averaged material properties for solving the thermal diffusion equation. That approximation will break down at higher frequencies, and solving the full heat diffusion equation becomes necessary for calculating the high frequency thermo-optic noise. The result for a $\rm Ta_2O_5:SiO_2$ coatings is shown in figure \ref{fig3}, again for a 19-doublet and a 8-doublet coating. Below 1~kHz the former agrees well with the thick-coating approximation, while the latter is better represented by the thin-coating approximation. Above about 10~kHz both approximations break down.
Finally, figure \ref{fig4} shows the thermo-optic noise of the two $\rm GaAs:Al_{0.92}Ga_{0.08}As$ coatings discussed in this paper. The cancellation effect that naturally occurs for the $18.5$ doublet coating results in a thermal noise at $100~{\rm Hz}$ fifty times below the one of the 40.5 doublet coating. By deviating from a simple $\lambda/4$ design such a cancellation effect can also be achieved for coating with higher reflectivities \cite{TaraThesis,RanaTBD}. 
The thick-coating approximation from \cite{Evans2008} provides good results for the 40.5 doublet coating in the gravitational-wave observation band below about 2~kHz. However none of the approximations is particularly useful for the 18.5 doublet, suggesting that estimating the extent to which a thermo-optic noise cancellation can be achieved requires the detailed heat flow analysis done in this paper.

\section{Conclusion}
\label{conclusion}
\We derived a coating correction factor for the photo-thermal transfer function in dielectric mirror coatings. \We showed that, depending on the depth at which the absorption in the coating occurs, this can lead to an enhancement and a sign flip of the photo-thermal transfer function at frequencies for which the diffusion length becomes comparable or smaller than the coating thickness. The thermo-optic transfer function for two mirror coatings important for the gravitational-wave community was calculated. The high frequency shape of the transfer function can be a powerful tool to distinguish between coating-intrinsic absorption and absorption on the mirror surface due to contamination. Another possible application is the use of the sign flip in the photo-thermal transfer function to stabilize radiation pressure feed-back in single carrier opto-mechanical systems. Finally the thermo-optic noise expression given by Evans et al. \cite{Evans2008} acquires additional corrections at high frequencies, and \we showed that estimating the quality of thermo-optic cancellation effects in crystalline coatings requires a detailed analysis of the heat flow.

\begin{acknowledgments}
\We would like to Rana Adhikari, Gregg Harry, Antonio Perreca and Matt Abernathy for fruitful discussions. This work was supported by the National Science Foundation grant PHY-1352511. This document has been assigned the LIGO Laboratory document number  LIGO-P1400212.
\end{acknowledgments}

\begin{table}[h]
\begin{tabular}{llrrl}
Parameters ${\rm Ta_2O_5\!:\!SiO_2}$  & Symbol                   & ${\rm SiO_2}$    & ${\rm Ta_2O_5}$     & Unit                   \\
\hline
Refractive Index (@1064 nm) & $n$                         & 1.45                      & 2.06 & -                      \\
Specific Heat                               & $C$                        & 746                       & 306 & J/kg/K                 \\
Density                                         & $\rho$                    & 2200                      & 6850 & kg/m${}^3$ \\
Thermal Conductivity                 & $\kappa$               & 1.38                      &  33  & W/m/K                  \\
Thermal expansion coef.           & $\alpha$                & 0.51                      & 3.6 & ppm/K                  \\
Thermo-Optic coef.  ($\rm 1{\mu}m$)& $\beta = \frac{dn}{dT}$ & 8             & 14 & ppm/K                  \\
Poisson ratio                                & $\sigma$               & 0.17                       &  0.23 & -                 \\
Young’s Modulus                        & $E$                        & 72.80                     & 140 & GPa
\end{tabular}
\caption{Parameters for fused silica (${\rm SiO_2}$) and tantulum-pentoxide (${\rm Ta_2O_5}$). The values are taken from \cite{Evans2008} and \cite{PhysRevD.70.082003}.  }
\label{SiO2}
\end{table}
\begin{table}[h]
\begin{tabular}{llrrl}
Parameter $\rm Al_{x}Ga_{1-x}As$ & Symbol                   & $x=0$    &  $x=0.92$ & Unit                   \\
\hline
Refractive Index (@1064 nm) & $n$                         &      3.48                 & 2.977 & -                      \\
Specific Heat                               & $C$                        &     330                  & 440 & J/kg/K                 \\
Density                                         & $\rho$                    &     5320                  & 3880 & kg/m${}^3$ \\
Thermal Conductivity                 & $\kappa$               &     55                 &  77  & W/m/K                  \\
Thermal expansion coef.           & $\alpha$                &      5.7               &  5.2 & ppm/K                  \\
Thermo-Optic coef. ($\rm 1{\mu}m$)& $\beta = \frac{dn}{dT}$ & 366           & 179 & ppm/K                  \\
Poisson ratio                                & $\sigma$               &       0.31           &  0.40 & -                 \\
Young’s Modulus                        & $E$                        &      85.3              & 83.6 & GPa
\end{tabular}
\caption{Parameters for a  $\rm GaAs\!:\!Al_{0.92}Ga_{0.08}As$ crystalline coating. The values are taken from \cite{Cole2013,Handbook}.  }
\label{AlGaAs}
\end{table}

\appendix
\section{Coating Reflectivity}
\label{App:Coat}
Here \we give a derivation for the partial derivatives of the coating reflectivity with respect to the round trip phase in coating layer $k$. 
For a dielectric stack with $N$ layers, each layer with index of refraction $n_k$, thickness $d_k$ and round trip phase $\phi_k=4 \pi n_k d_k / \lambda_0$, \we can define right- and left-travelling modes $\Psi^{\rm R}$ and $\Psi^{\rm L}$ at every interface.
\We assume the light is incident from the left towards the coating at $z=0$, and find the transfer matrix relation
\begin{equation}
\left(\begin{array}{c} \Psi^{\rm R} \\ \Psi^{\rm L} \end{array} \right)_{k+1} = Q_k D_k \left(\begin{array}{c} \Psi^{\rm R} \\ \Psi^{\rm L} \end{array} \right)_i ,
\end{equation}
where
\begin{equation}
D_k= \left( 
\begin{array}{cc}
\e^{-i \phi_k /2}& 
0 \\ 
0 & 
\e^{i \phi_k /2} 
\end{array} \right)
\end{equation}
is the propagator through the layer and
\begin{equation}
Q_k= \frac{1}{2 n_{k+1}}\left( 
\begin{array}{cc}
n_{k+1}  + n_k & 
n_{k+1}  - n_k \\ 
n_{k+1}  - n_k & 
n_{k+1}  + n_k 
\end{array} \right)
\end{equation}
is the transition matrix from layer $k$ to layer $k+1$.
The transfer matrix for the total coating is
\begin{equation}
M = Q_N D_N ... Q_k D_k  ... Q_1 D_1 Q_0 ,
\end{equation}
with $Q_0$ the transition matrix from vacuum to layer 1.
$M$ is related to coating reflectivity $r$ and transmission $t$ by
\begin{equation}
\label{eq:Mr}
M \left( 
\begin{array}{cc} 1 \\ r \end{array} \right)= \left( 
\begin{array}{cc} t \\ 0 \end{array} \right) ,
\end{equation}
which can easily be solved for the reflectivity $r$.
\We will need the derivative of $r$ with respect to the round trip phase $\phi_k$ in layer $k$. Thus \we also want 
\begin{equation}
\frac{\partial M}{\partial \phi_k} = Q_N D_N ... Q_k D_k \!\!\left( \!\!\!
\begin{array}{cc} -i/2 & 0 \\  0 & i/2 \end{array}
\!\!\! \right)\!\!Q_{k-1} D_{k-1}  ... Q_1 D_1 Q_0 .
\end{equation}
Using the chain rule on \eq{eq:Mr} \we find
\begin{equation}
{\frac{1}{r}\frac{\partial r}{\partial \phi_{k}}} = 
{\frac{1}{M_{21}}\frac{\partial M_{21}}{\partial \phi_{k}}} - 
{\frac{1}{M_{22}}\frac{\partial M_{22}}{\partial \phi_{k}}} ,
\end{equation}
which can be used directly in \eq{eq:dphidphi}, and we find
\begin{equation}
\label{eq:parder}
\frac{\partial \phi_{\rm c}}{\partial \phi_k} = {\rm Im} \left(
{\frac{1}{M_{21}}\frac{\partial M_{21}}{\partial \phi_{k}}} - 
{\frac{1}{M_{22}}\frac{\partial M_{22}}{\partial \phi_{k}}}   \right) .
\end{equation}

\section{Coating Heat Diffusion}
\label{App:CoatDiffuse}
Here \we give the solution to the one-dimensional heat diffusion equation $\xi_k^2 T = T'' $ across the whole coating and substrate. As discussed in the main text \we assume that the heat $j$ is deposited on the first interface layer.

The boundary conditions require that 
 $T$ and $j=-\kappa_k \nabla T$ are continuous everywhere.
In each layer we have $\xi_k=\sqrt{{i \omg C_k \rho_k}/{\kappa_k}}$, and the solution has the form given in \eq{eq:layersolution}.
At $z=z_0$ the heat flow $j$ and temperature $T$ are related to $T_R$ and $T_L$ via
\begin{equation}
\left( \begin{array}{c} j \\ T \end{array} \right) 
=
\left( \begin{array}{cc} {\kappa \xi} &   \\   & 1  \end{array} \right) 
\left( \begin{array}{cc} 1 & -1 \\ 1 & 1  \end{array} \right)
\left( \begin{array}{c} T_R \\ T_L  \end{array} \right) .
\end{equation}
\We can therefore define
\begin{equation}
E_k =
\left( \begin{array}{cc} {\kappa_k \xi_k} &   \\   & 1  \end{array} \right) 
\left( \begin{array}{cc} 1 & -1 \\ 1 & 1  \end{array} \right)
\left( \begin{array}{cc} \e^{-\xi_k \frac{d_k}{2}} &   \\   & \e^{\xi_k \frac{d_k}{2}}  \end{array} \right)
\end{equation}
and
\begin{equation}
F_k =
\left( \begin{array}{cc} \e^{-\xi_k \frac{d_k}{2}} &   \\   & \e^{\xi_k \frac{d_k}{2}}  \end{array} \right)
 \frac{1}{2} \left( \begin{array}{cc} 1 & 1 \\ -1 & 1  \end{array} \right)
\left( \begin{array}{cc} \frac{1}{\kappa_k \xi_k} &   \\   & 1  \end{array} \right) .
\end{equation}
The operator $D_k=E_k F_k$ propagates the heat flow and temperature field by across the layer $k$:
\begin{equation}
\left( \begin{array}{c} j \\ T \end{array} \right)_{k,k+1} 
= D_k
\left( \begin{array}{c} j \\ T \end{array} \right)_{k-1,k} ,
\end{equation}
while $F_k$ reads out the temperature in the middle of the coating:
\begin{equation}
\left( \begin{array}{c} T_R \\ T_L  \end{array} \right)_{k,{\rm middle}}
= F_k
\left( \begin{array}{c} j \\ T \end{array} \right)_{k-1,k} .
\end{equation}
To fulfill the global boundary conditions, \we define for the substrate
\begin{equation}
F_s =
 \frac{1}{2} \left( \begin{array}{cc} 1 & 1 \\ -1 & 1  \end{array} \right)
\left( \begin{array}{cc} \frac{1}{\xi_s \kappa_s} &   \\   & 1  \end{array} \right)
\end{equation}
and
\begin{equation}
M =   F_s D_N ... D_2 D_1 ,
\end{equation}
which fulfills
\begin{equation}
\label{eq:M2}
M \left( \begin{array}{c} j \\ T \end{array} \right)_{0,1} 
= \left( \begin{array}{c} T_R \\ 0  \end{array} \right)_{N,s} ,
\end{equation}
where the left-propagating mode is set to zero in the substrate to keep the temperature finite at plus infinity. 

This can be solved for $T_{0,1}$. The temperature in the middle of each coating layer $i$ becomes
\begin{equation}
\label{eq:tempprofile}
\left( \begin{array}{c} T_R \\ T_L  \end{array} \right)_{k,{\rm middle}} = F_k D_{k-1} ... D_1 
\left( \begin{array}{c} 1 \\ -\frac{M_{21}}{M_{22}} \end{array} \right)  j .
\end{equation}
Similarly equation~\ref{eq:M2} directly gives us $T_{R,N,s}$. Both results can now be used in equations \ref{eq:Tidi} and \ref{eq:Tsds}.

The matrix formalism discussed here can also be extended to bulk heating. For this \we use 3x3 matrices with the third row equal to (0,0,1). The field vectors are also extended to 
\begin{equation}
\left( \begin{array}{c} j \\ T \\1 \end{array} \right) 
=
\left( \begin{array}{ccc} \kappa \xi & -\kappa \xi & \\ 1 & 1 & \\ & & 1  \end{array} \right)
\left( \begin{array}{c} T_R \\ T_L \\ 1 \end{array} \right) .
\end{equation}
The layer propagation matrices $D_k$ become
\begin{equation}
D_k \!=\!\!\!
\left(\!\!\! \begin{array}{ccc} \cosh {\xi_k d_k} &\!\!\!\!\! -\kappa_k \xi_k  \sinh {\xi_k d_k}&\!\!\!\!\!  \frac{p_k}{\xi_k}\sinh {\xi_k d_k}\\
 -\frac{1}{\kappa_k \xi_k}  \sinh {\xi_k d_k} &\!\!\!\!\! \cosh {\xi_k d_k} &\!\!\!\!\! -\frac{2p_k}{\kappa_k \xi_k^2} \sinh^2 \frac{\xi_k d_k}{2}\\
 & &\!\!\!\!\! 1  \end{array} \right) 
\end{equation}
where $p_k$ is the bulk heating power density in layer $k$. For substrate heating the boundary conditions deserve some attention. Since it is non-adiabatic, the heat flow $j$ should asymptote to zero, whereas the temperature will asymptote to the adiabatic value $T_{adi}=q_s/(\xi_s^2 \kappa_s)$. This extension was used for the thermal noise calculation in appendix \ref{App:Noise}.

\section{Noise calculation}
\label{App:Noise}
For the thermo-optic noise calculation \we follow \cite{Levin2008}. According to the discussion in section \ref{Derive} the displacement fluctuations of the mirror as seen by a laser beam are given by
\begin{equation}
\label{eq:dphic3}
\del z =  \int\limits_0^\infty dz  \int d^2 r  q(z) q(r) \delta T(r,z) ,
\end{equation}
with the readout functions
\begin{equation}
q(r) = \frac{2}{\pi \w^2} \e^{-2\frac{r^2}{\w^2}}
\end{equation}
normalized to $\int d^2 q(r)=1$, and
\begin{equation}
q(z) = \left[ \frac{\partial \phi_{\rm c}}{\partial \phi_k} (\beta_k \!+\! \bar{\alpha}_k n_k) 
\!+\!  \bar{\alpha}_k  \right]_{k(z)} ,
\end{equation}
where the bracket is evaluated for the corresponding coating layer. In time domain the coating is now heated with the energy density
\begin{equation}
\frac{dQ}{dV} = T ds = T F_0 \cos{(\omega t)} q(z) q(r) ) ,
\end{equation}
where $s$ is the entropy per unit volume. $F_0$ is the entropy drive amplitude as introduced by Levin \cite{Levin2008}. It will cancel in the final expression \ref{eq:MasterNoise}.
 Switching back to frequency domain we find for the heating power per volume, $p$:
\begin{equation}
\label{eq:BulkHeat}
p = i \omega \frac{dQ}{dV} = i \omega T F_0 q(z) q(r).
\end{equation}

The cycle-averaged dissipated power is
\begin{equation}
W_{\rm diss} =\frac{1}{T} \int dV \frac{|j(z,r)|^2}{2 \kappa}
\end{equation}
The factor of 2 in the denominator  is required because we are working in Foruirer domain and $j(z,r)$ is complex. Since we again neglect radial diffusion, the radial dependence reduces to 
\begin{equation}
\int d^2r q^2(r) = \frac{1}{\pi \w^2}, 
\end{equation}
and we find for the cycle-averaged dissipated power
\begin{equation}
W_{\rm diss} =\frac{1}{2 T \pi \w^2} \int\limits_0^\infty \frac{|j(z)|^2}{\kappa} dz.
\end{equation}
Given the bulk heating $p$ from equation \ref{eq:BulkHeat} as input, 
we can use the approach layed out in appendix \ref{App:CoatDiffuse} to calculate the last integral. 

Finally, directly following Levin's approach \cite{Levin2008}, the thermo-optic power spectral density for the readout degree of freedom $\del z$ is then given by
\begin{equation}
\label{eq:MasterNoise}
S_{\delta T}(f) = \frac{8 k_B T}{\omega^2} \frac{W_{\rm diss}}{F_0^2} .
\end{equation}

\bibliography{LongIFO}
\bibliographystyle{plain}
\end{document}